\documentclass[twocolumn,prl,amsmath,amssymb,showpacs,superscriptaddress,floatfix]{revtex4}
\usepackage{graphicx}
\usepackage{bm}
\usepackage{lineno,hyperref}

\usepackage{epsf}
\usepackage{epstopdf}
\begin{document}
\title{Quasicrystal vs Glass Transition: comparison structural and dynamical properties.}

\author{Yu. D. Fomin}
\affiliation{ Institute for High Pressure Physics RAS, 108840
Kaluzhskoe shosse, 14, Troitsk, Moscow, Russia}
\affiliation{Moscow Institute of Physics and Technology, 9 Institutskiy Lane, Dolgoprudny City, Moscow Region, Russia }

\date{\today}

\begin{abstract}
Quasicrystals are solid structures with symmetry forbidden by crystallographic rules. Because of this some
structural characteristics of quasicrystals, for instance, radial distribution function, can look similar to the ones of amorphous phases.
This is of principal importance since radial distribution function is the main property to characterize the structure in
molecular simulation. In the present paper we compare the radial distribution functions and dynamical properties
of three systems in the vicinity of glass transition, quasicrystal formation and crystallization. We show
that in spite of similarity of radial distribution functions the dynamical properties of a system
in the vicinity of quasicrystal are qualitatively equivalent to the ones of crystal. Because of this
combination the radial distribution functions with investigation of dynamics of the liquid allows
unambiguously distinguish glass and quasicrystal.

\end{abstract}


\maketitle

All substances except helium crystallize upon cooling. The crystalline phases are characterized by
strict long-range order both translational and orientational. However, if the cooling proceeds
very fast the crystallization can be avoided and the system transforms into glassy state. Glass
transition is among the most complex problems of condensed matter physics. Until now it is not
clear whether it is a thermodynamic transition or purely kinetic effect. Usually the glass transition
temperature is defined as the temperature at which the viscosity of the substance becomes as high
as $10^{12}$ $Pa \cdot s$ \cite{kobbinder}. Glassy state demonstrates some features of solids and some features of liquids.
Among the solid-like features the most important is shear rigidity. At the same time the diffraction pattern of
glasses does not demonstrate any peaks, i.e. it looks liquid-like.

The wealth of solid structures is not restricted by crystals and glasses only. Another type of solids
is quasicrystals (QC). QCs are ordered structures with symmetry restricted by crystallographic rules.
Although initially QCs were discovered in metallic alloys \cite{qc-discovery}, later on
it was found that they can be formed in other systems too, for instance, in water solution of micelles \cite{qc-micelles},
graphene bilayers \cite{qc-graphene}, etc. Moreover, it was found that even one component systems demonstrate quasicrystalline phase in
computer simulation in theree (see, for instance, \cite{qc1,qc2,qc3,qc4}) and two dimensional (see, for instance, \cite{qc2d1,qc2d2,qc2d3,qc2d4}) spaces.

Several studies report formation of QCs in so-called core-softened systems, i.e. the systems with
softening of repulsive core of the interaction potential. QCs were observed in such models as
Dzugutov potential \cite{qc1}, Lennard-Jones plus Gauss model \cite{qc3}, repulsive shoulder system (RSS) \cite{ryltsev,ryltsev1} and some other models.
Interestingly, in Ref. \cite{ryltsev} it was found that the ability of a liquid to transform into quasicrystal can be predicted
basing on its radial distribution function (rdf) $g(r)$. This assumption was validated by simulation
of three different systems which form quasicrystalline phase. In the later works the same group of
authors found more different quasicrystalline phases in RSS with different parameters of the interaction potentials were discovered \cite{ryltsev2}.

\begin{figure}
\includegraphics[width=8cm,height=8cm]{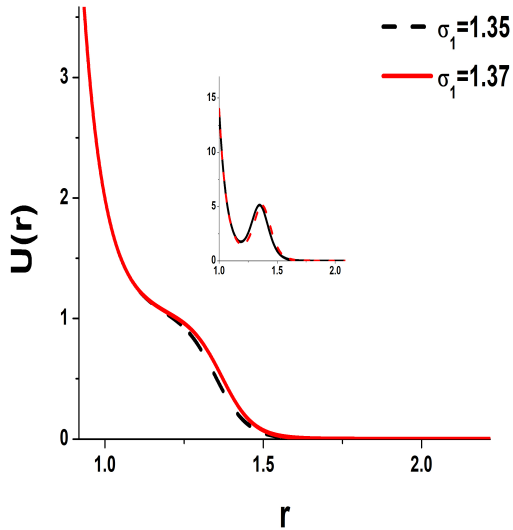}%

\caption{\label{fig-pot} The potential of Repulsive Shoulder System (RSS) for two values of $\sigma_1$: 1.35 and 1.37.
The inset shown the force in the system $F=-\partial U / \partial r$.}
\end{figure}

The phase diagram of RSS was widely investigated in a set of papers \cite{we1,we2,we3,we4}. This system is defined
by the interaction potential of the form:

\begin{equation}\label{pot}
U(r)/ \varepsilon = \left( \frac{\sigma}{r}
\right)^{14}+0.5\left(1- tanh(k(r-\sigma_1)) \right).
\end{equation}
In our previous studies the parameter $k$ was set to $k=10.0$. The potential for $k=10$ and $\sigma_1=1.35$ and $1.37$
 is shown in Fig. \ref{fig-pot}. The parameter $\sigma_1$ determines the width of
the repulsive shoulder. It is convenient to express all quantities
in the units of the potential, i.e. the parameter $\varepsilon$ serves
as a unit of energy and $\sigma$ as a unit of length. All other quantities
can be expressed from these parameter. Below all quantities are given in these
reduced units.

The phase diagram of RSS demonstrates extremely strong dependence on the parameters of the potential. Moreover, it demonstrates
many anomalous features similar to the anomalies of water \cite{we2,we3,we5,we6,we7,we8,we9,we10,we11}.
Already in the first paper on RSS \cite{we1} existence of glass transition in the system with $\sigma_1=1.35$ was found.
Later on more elaborate study of this glass transition was performed in \cite{ryltsev-glass}. It was shown
that in the range of densities from about $0.5$ up to about $0.75$ the system demonstrates the glass transition.
Interestingly, if the width of repulsive shoulder is slightly changed to $\sigma_1=1.37$ then a QC phase is formed instead of
glass transition \cite{ryltsev} which once again shows the extreme sensitivity of the phase diagram of RSS to the parameters of the potential.

Having established such interplay between the glass transition and QC formation it becomes of great importance to
distinguish between these two phases. In computer simulation the structure of the system is typically characterised by
rdfs. Interestingly, the rdfs of QC do not demonstrate any strict ordering (see, for instance, Fig. 1b of Ref. \cite{ryltsev})
and look very similar to the ones of glass. Because of this the quasicrystalline structure can be erroneously
classified as glass. Correct identification of glass or quasicrystall requires more elaborate study of structural properties
of the system, such as Steinhard-Nelson order parameters \cite{sn,sn1} or diffraction patterns.

In the present paper we show that the dynamical properties of the system are very different in the case of QC formation
and glass transition. It means that they can be used to distinguish between them even without calculation
of bond-order parameter or diffraction patterns. Importantly, the dynamical properties such as means square
displacement (MSD) and intermediate scattering function are typically calculated in simulations of glass transition. Therefore,
no additional calculations such as diffraction pattern or order parameters is required.

In the present study we simulate by means of molecular dynamics method a system of 4000 particles interacting with
RSS potential with $\sigma_1=1.35$ for glass-forming system and with $\sigma_1=1.37$ for the system forming QC.
Cubic box with periodic boundaries is used in both cases. The time step is set to $dt=0.001$ for $\sigma_1=1.35$ and $dt=0.01$ for $\sigma_1=1.37$.
In both cases $1 \cdot 10^8$ steps are made for equilibration of the system. NVT ensemble (constant number of particles N, volume V and temperature T) is
used at this stage. After that microcanonical simulation (constant number of particles N, volume V and internal energy E)
for more $5 \cdot 10^7$ steps is performed. We calculate the internal energy and pressure of the system. To characterise the
structure we compute the radial distribution functions and the structure factors of the system. The dynamical
properties are characterized by mean square displacement, the intermediate scattering function $F_s$ and stress-stress autocorrelation function.

The intermediate scattering function if defined as $F_s(\bf{k},t)=\sum_{i=1}^N e^{-i \bf{kr}_i}$, where the wave
vector $k$ is selected as the first maximum of the structure factor. 
The stress tensor is defined as $\sigma_{xy}=\sum_{i=1}^N m_iv_{i,x}v_{i,y}+1/2 \sum_{i \neq j} x_{ij}F_{y,ij}$,
where $v_{i,x}$ is x component of velocity of $i-$th particle, $x_{ij}=x_i-x_j$ the x component of the vector connecting
i-th and j-th particles and $F_{y,ij}$ is y component of the force between these particles. The off-diagonal components of the stress
tensor can be used to calculate the shear viscosity via Green-Kubo relation
$\eta = \frac{V}{k_BT} \int_{0}^{\infty} \sigma_{xy}(t)\sigma_{xy}(0) dt$.

In the case of RSS with $\sigma_1=1.35$ the density is $\rho=0.53$ where the system demonstrates glass transition at low
temperature \cite{we1,ryltsev-glass}. Additional calculations are made  for the density $\rho=0.4$ where the system crystallizes upon cooling.

RSS with $\sigma_1=1.37$ is simulated at $\rho=0.474$ where appearance of QC phase is found \cite{ryltsev}.

\begin{figure}
\includegraphics[width=8cm,height=8cm]{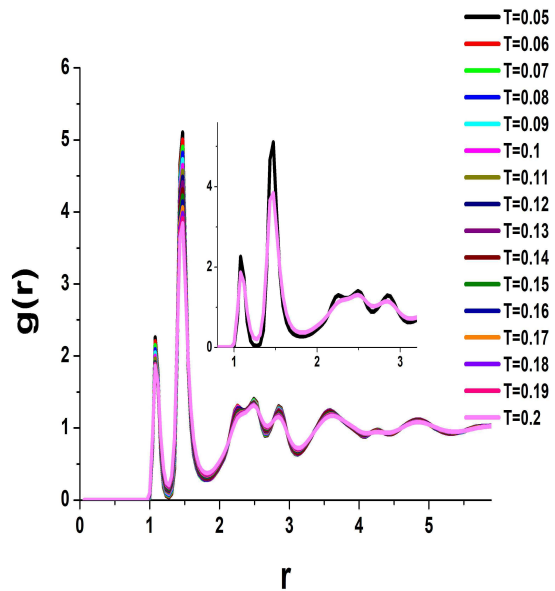}%

\caption{\label{rdf-gl} Radial distribution functions of RSS with $\sigma_1=1.35$ at $\rho=0.53$. This system
experiences the glass transition. The inset enlarges the highest and the smallest temperature.}
\end{figure}

Fig. \ref{rdf-gl} shows rdfs of RSS with $\sigma_1=1.35$ at $\rho=0.53$, i.e. in the glass forming region. The temperature is
from $T_{max}=0.2$ and to $T_{min}=0.05$. One can see that the first peak of rdf splits into two subpeaks which is
an intrinsic property of systems with two length scales. More peaks appear at low temperatures, however, these peaks are not
characteristic to any crystalline structure and they signalize that the structure is frozen in some glassy state.

\begin{figure}
\includegraphics[width=8cm,height=8cm]{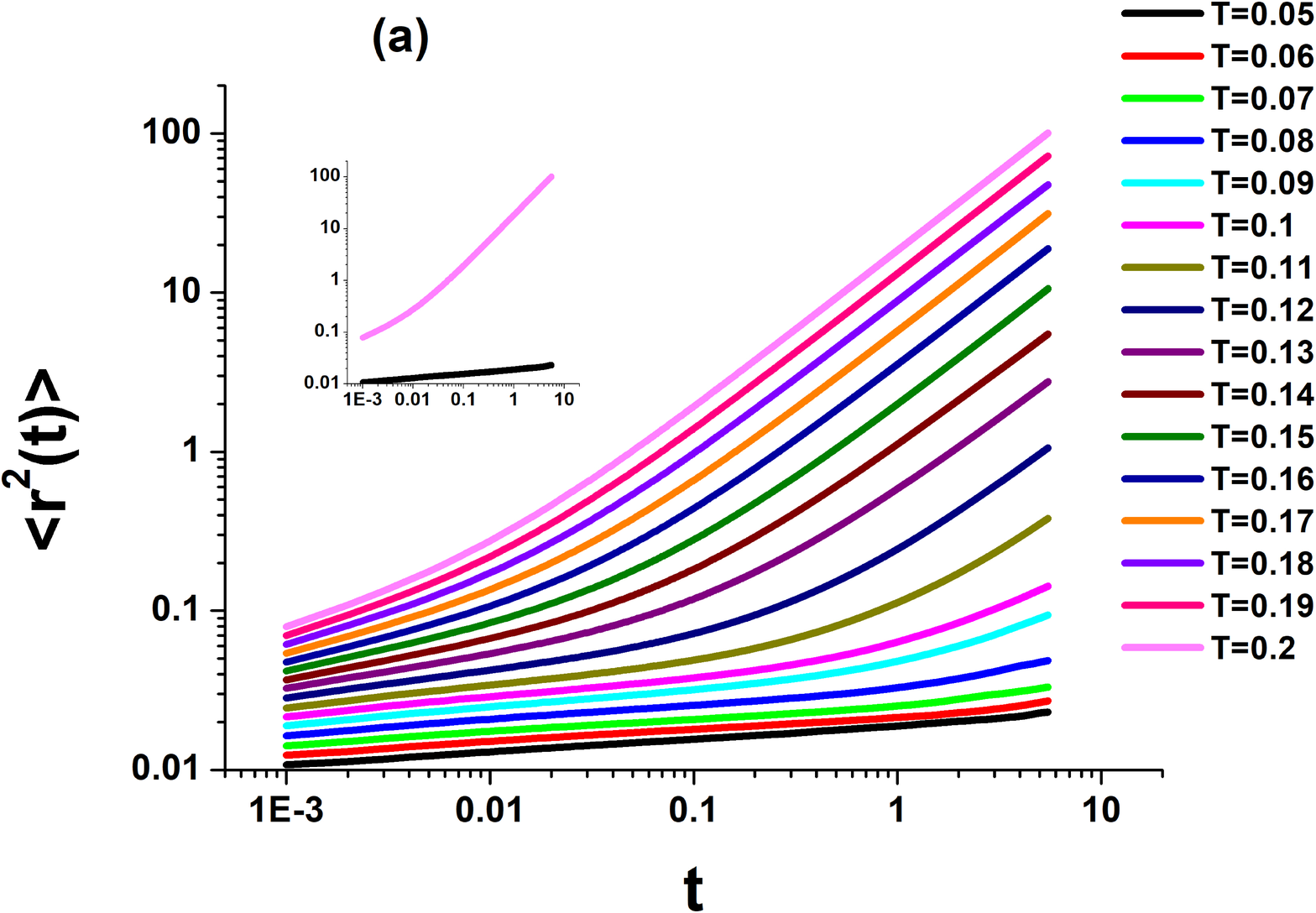}%

\includegraphics[width=8cm,height=8cm]{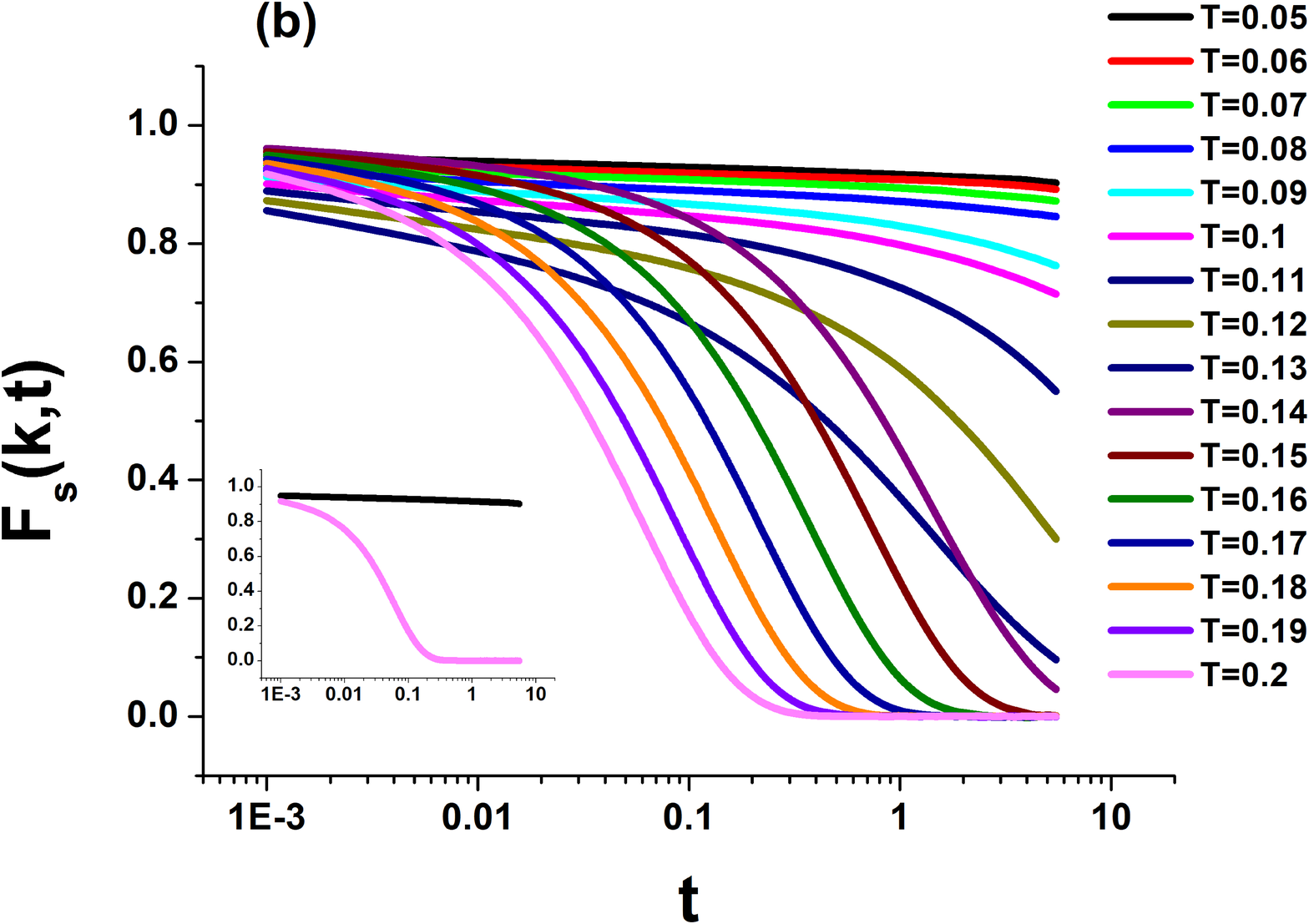}%

\caption{\label{msd-gl} (a) Mean square displacement and (b) intermediate scattering function at $k=5.44$ of RSS with $\sigma_1=1.35$ at $\rho=0.53$. This system
experiences the glass transition. The inset enlarges the highest and the smallest temperature. The insets in both panels
enlarge the highest and the smallest temperature.}
\end{figure}

The appearance of the glassy state is confirmed by the calculations of MSD and intermediate scattering function (Fig. \ref{msd-gl} (a) and (b)).
One can see rapid decay of $F_s(k,t)$ at high temperatures. As the temperature is lowered the decay becomes much slower
and finally the intermediate scattering function does not decay noticeably within the simulation time. Analogously,
MSD rapidly increases at high temperature, whilst at low temperatures the particles do not leave their cages: the MSD
does not exceed 0.05 particle diameters within the simulation time.

\begin{figure}
\includegraphics[width=8cm,height=8cm]{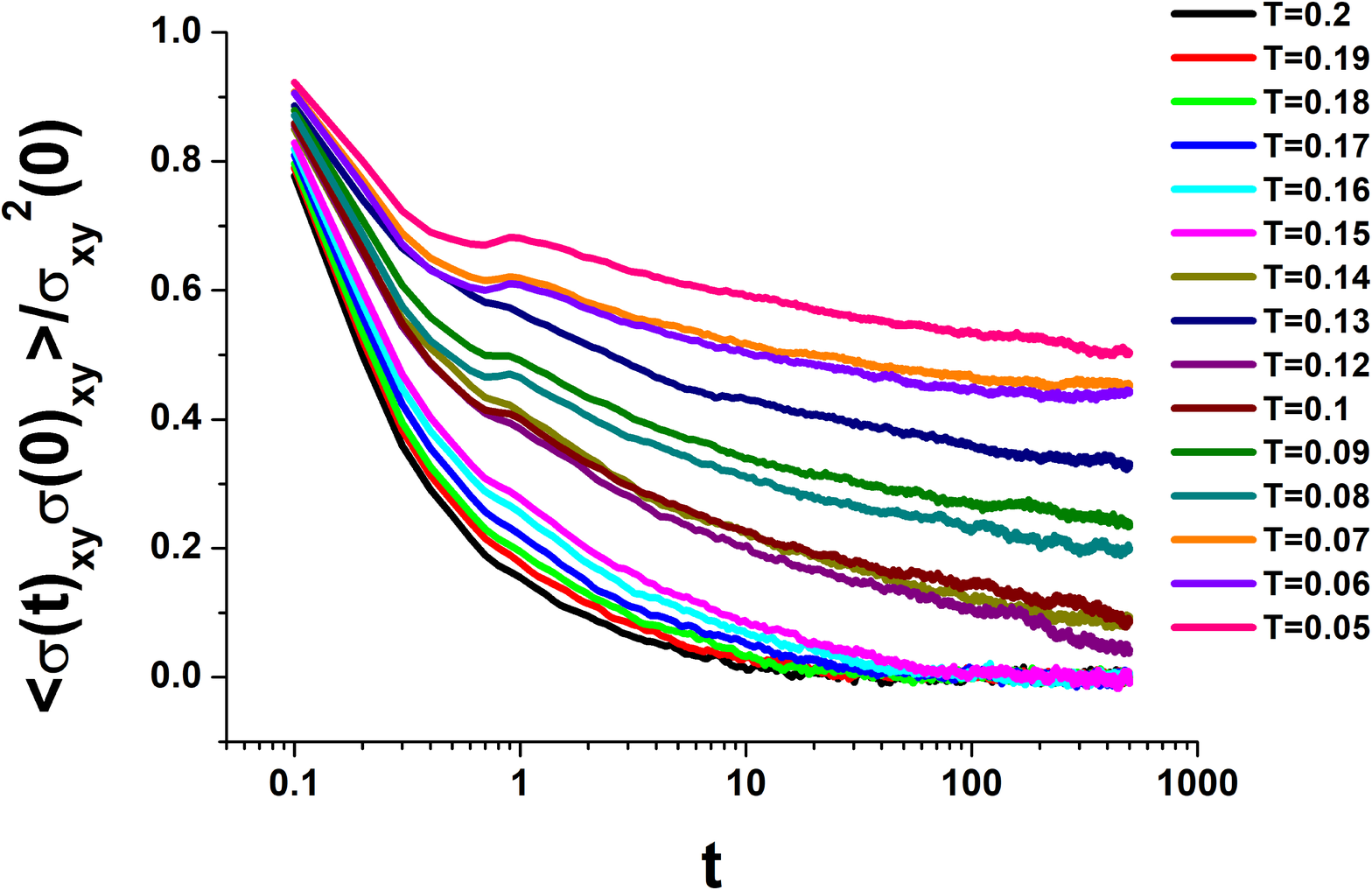}%

\caption{\label{visc-gl} Stress autocorrelation function for RSS with $\sigma_1=1.35$ at $\rho=0.53$.}
\end{figure}

Fig. \ref{visc-gl} shows the stress autocorrelation function of the same system. At high temperature the stress autocorrelation
decays to zero. However, at $T=0.12$ it does not reach zero within the simulation time, and therefore the shear viscosity
cannot be obtained by our calculations for the temperatures $T \leq 0.12$. Moreover, the time of decay of stress autocorrelation functions
becomes larger with lowering of the temperature, which means that the system becomes more viscous. This conclusion can be made
without precise calculation of the viscosity which becomes impossible in frames of the present work.

\begin{figure}
\includegraphics[width=8cm,height=8cm]{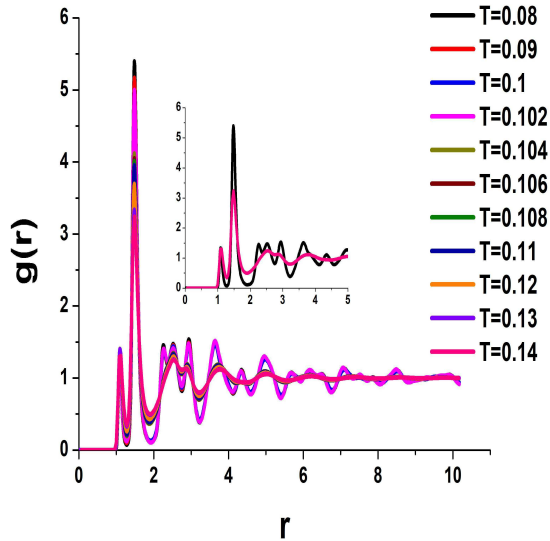}%

\caption{\label{rdf-qc} Radial distribution functions for RSS with $\sigma_1=1.37$ at $\rho=0.474$. The inset enlarges the rdfs for
the highest and the smallest temperatures.}
\end{figure}

\begin{figure}
\includegraphics[width=8cm,height=8cm]{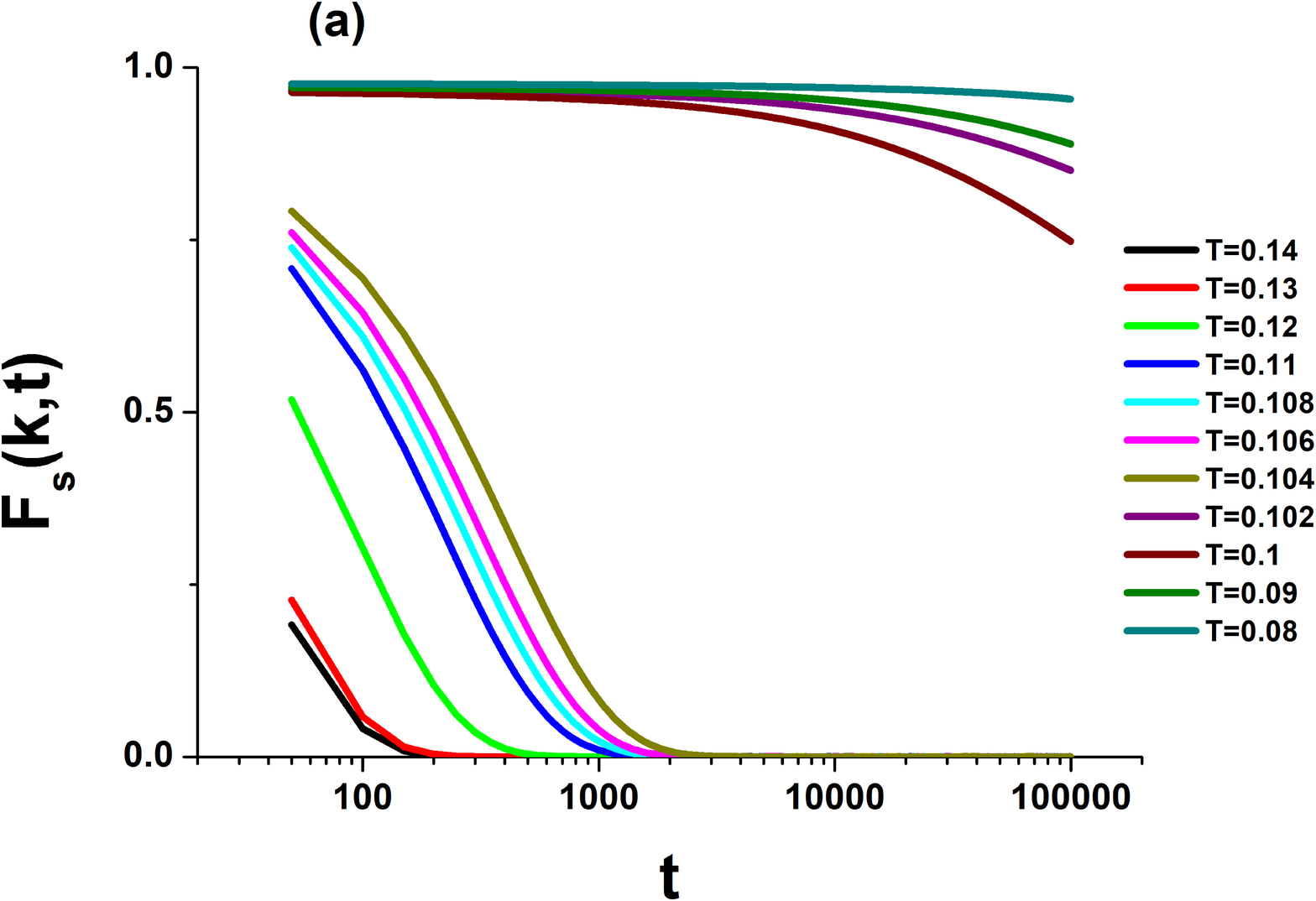}%

\includegraphics[width=8cm,height=8cm]{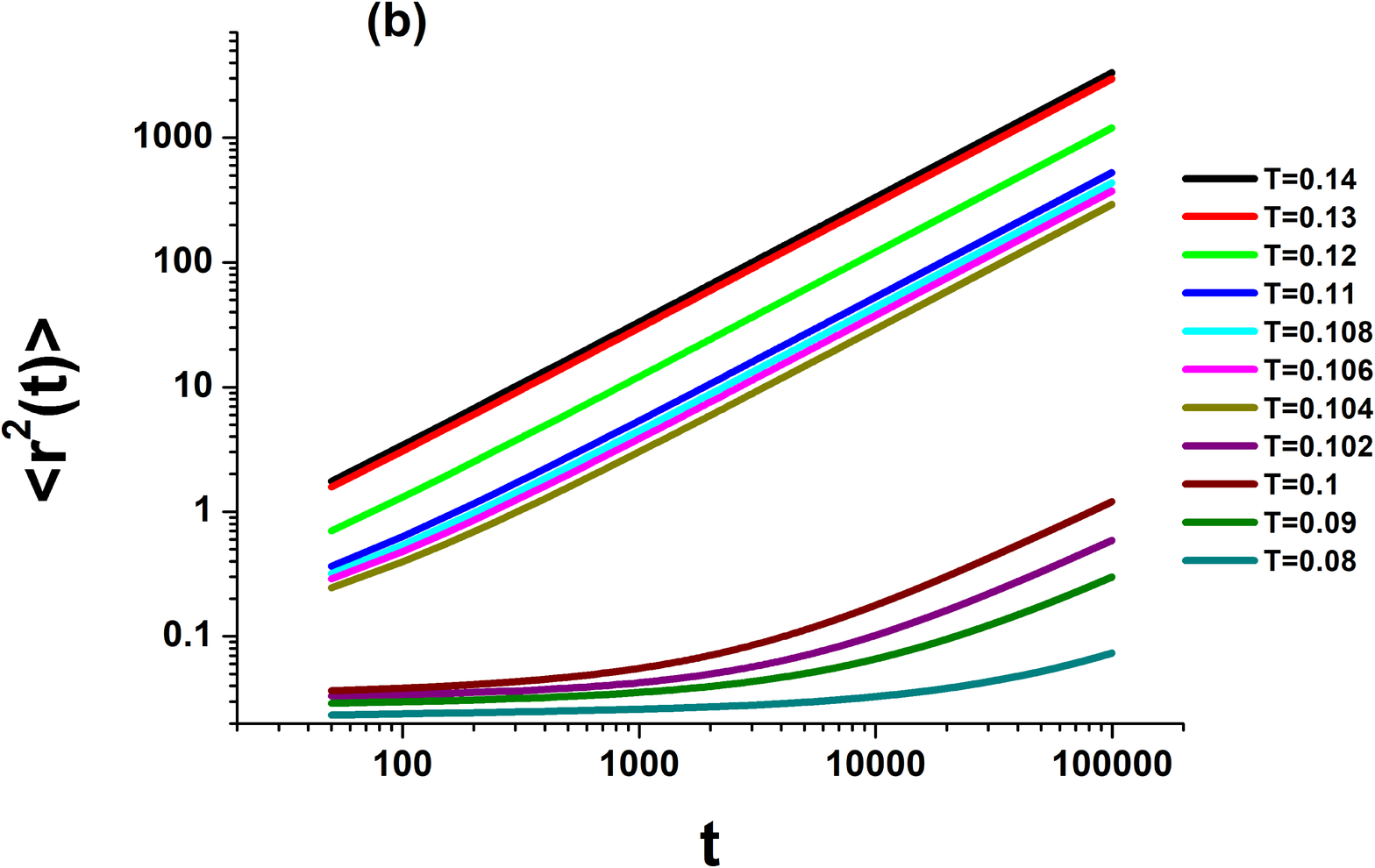}%

\caption{\label{msd-qc} (a) Intermediate scattering function at $k=5.24$ and (b) mean square displacement of RSS with $\sigma_1=1.37$ at $\rho=0.474$. }
\end{figure}

The situation becomes different in the case of the formation of QC. Fig. \ref{rdf-qc} shows the rdf of the RSS with $\sigma_1=1.37$
at $\rho=0.474$ at temperature from $T=0.08$ up to $T=0.14$. The formation of QC takes place at $T \leq 0.102$. The rdfs of the
system demonstrate some peaks and looks relatively ordered. However, they are not much different from the ones of glass and
indeed may be considered as very low temperature glass. However, the behavior of intermediate scattering function and
MSD is extremely different from the one of glass forming system. Fig. \ref{msd-qc} (a) and (b) show the
intermediate scattering functions and the MSD of the RSS with $\sigma_1=1.37$. One can see that at high temperature
they looks like the ones of a normal liquid. However, when the temperature is slightly change from $T_1=0.104$ to
$T_2=0.102$ the curves of both $F_s$ and MSD experience sharp qualitative change: below $T=0.102$ $F_s$ stops to decay
and MSD stops to increase. Importantly, the in the case of QC formation one observes a sharp change of the behavior, while
in the case of glass formation all quantities change smoothly.

\begin{figure}
\includegraphics[width=8cm,height=8cm]{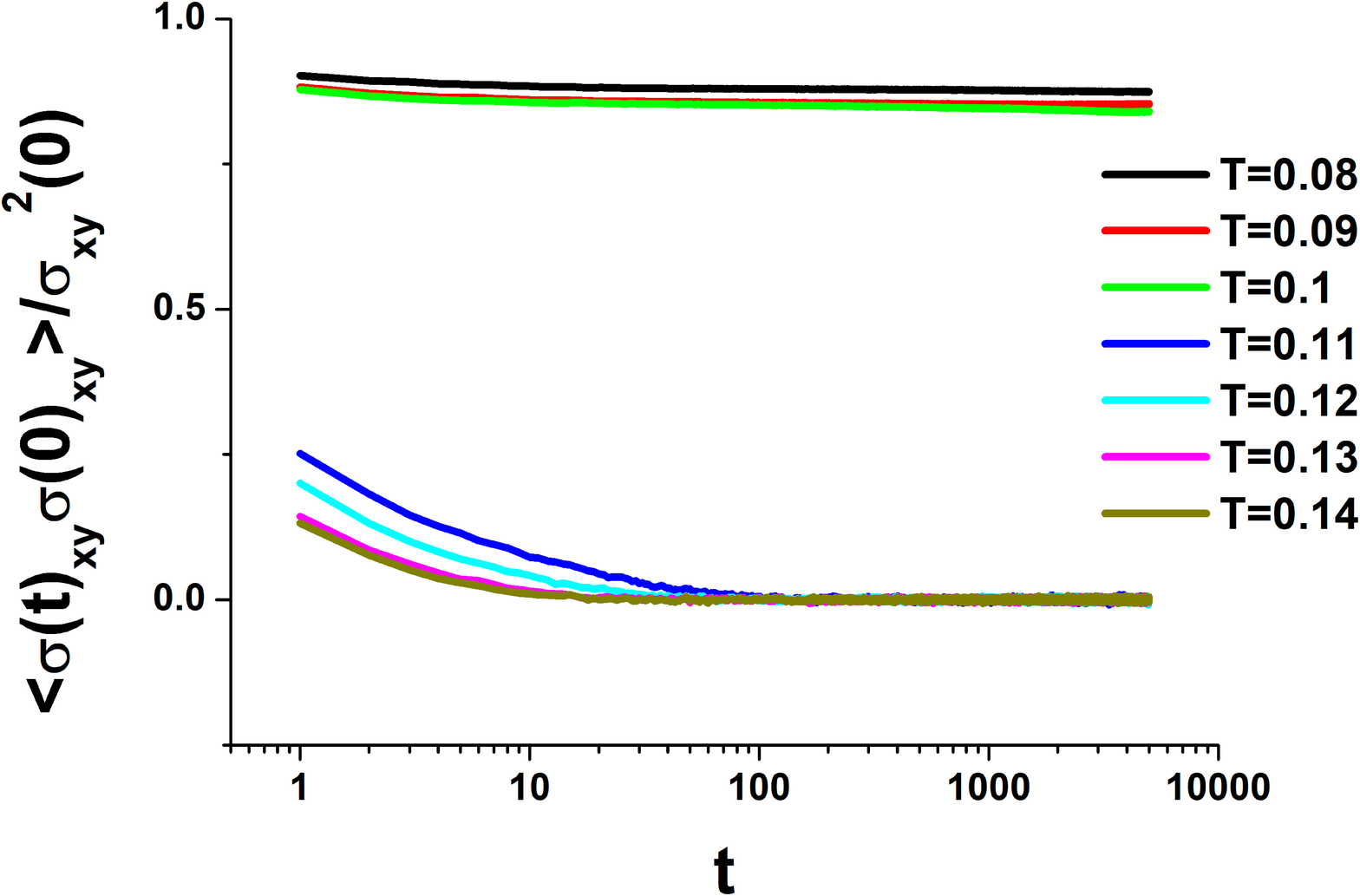}%

\caption{\label{visc-qc} Stress autocorrelation function for RSS with $\sigma_1=1.37$ at $\rho=0.474$.}
\end{figure}

The jump-like nature of the dynamical properties becomes even more evident from the behavior of stress autocorrelation
function (Fig. \ref{visc-qc}). One can see that the stress autocorrelation functions do not demonstrate any decay below
the point of QC formation, i.e. formally the viscosity becomes infinite, which corresponds to the solid state of matter.

\begin{figure}
\includegraphics[width=8cm,height=8cm]{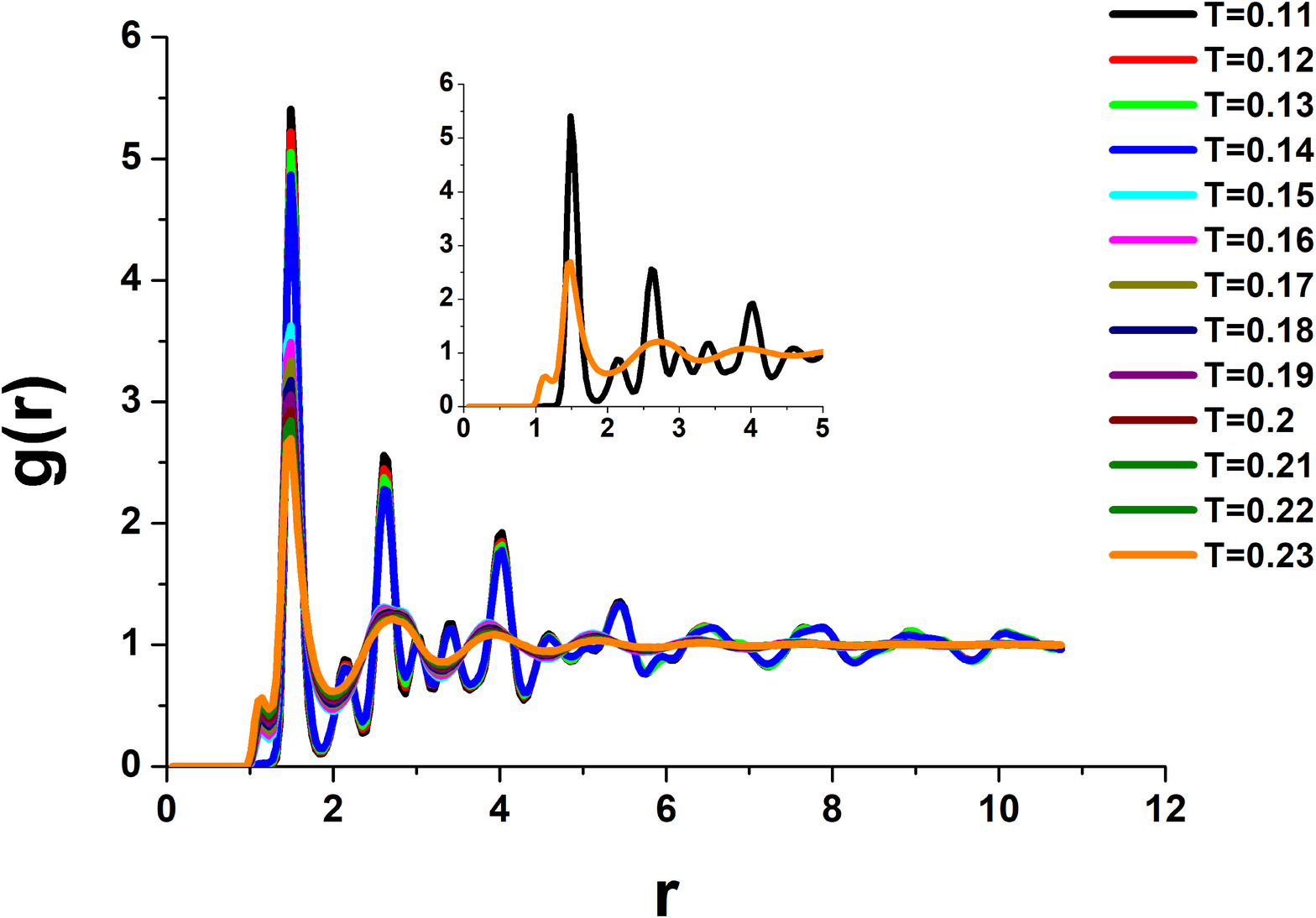}%

\caption{\label{rdf-cr} Radial distribution functions for RSS with $\sigma_1=1.35$ at $\rho=0.4$. The inset enlarges the rdfs for
the highest and the smallest temperatures.}
\end{figure}

\begin{figure}
\includegraphics[width=8cm,height=8cm]{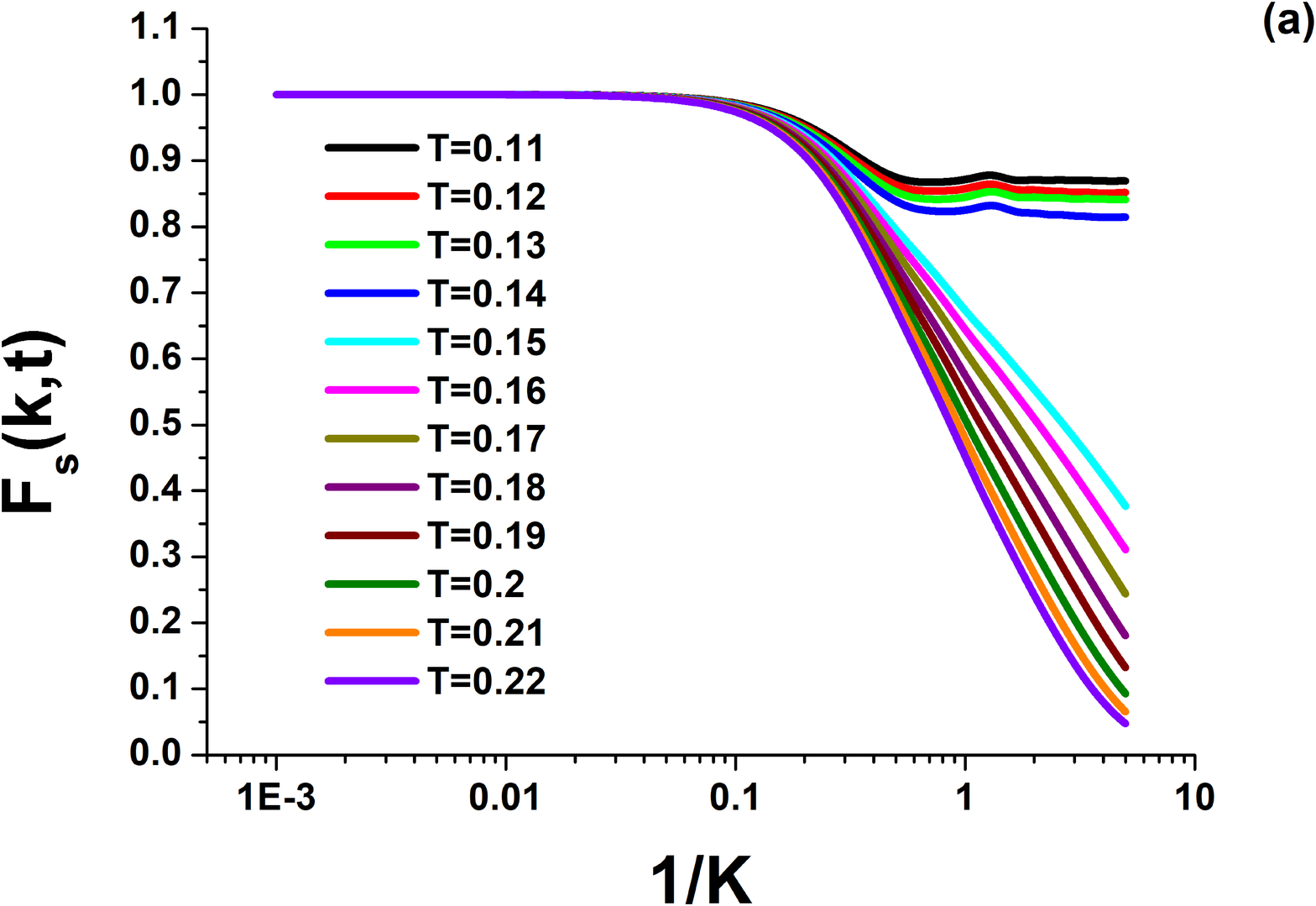}%

\includegraphics[width=8cm,height=8cm]{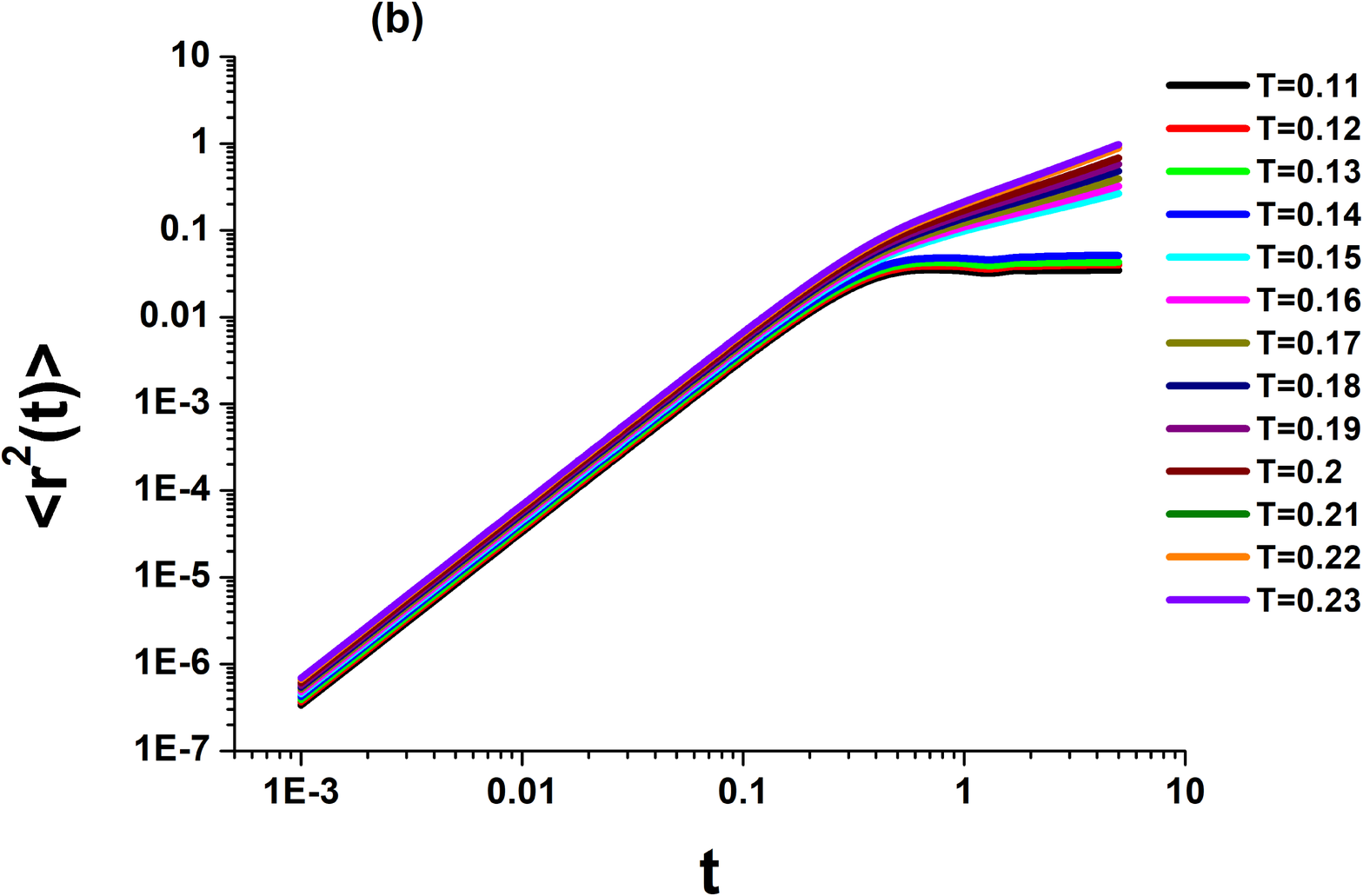}%

\caption{\label{msd-cr} (a) Intermediate scattering function at $k=4.96$  and (b) mean square displacement of RSS with $\sigma_1=1.35$ at $\rho=0.4$.}
\end{figure}

We compare the behavior of QC forming system with the one of the system which forms a crystal. For this comparison
we choose the RSS with $\sigma_1=1.35$ at the density $\rho=0.4$ where the system transforms into face centered cubic (FCC) lattice under cooling.
Fig. \ref{rdf-cr} shows the rdfs of the system. One can see that at high temperature the system is liquid, while at low temperature
numerous ordered peaks are observed, which means that the system crystallizes. Fig. \ref{msd-cr} (a) and (b) demonstrate
the intermediate scattering function and MSD of the system at different temperatures. The qualitative behavior of these quantities is
the same that in the case of QC formation: upon a tiny change of the temperature (from $T_1=0.15$ to $T_2=0.14$) the intermediate scattering
function stops to decay, while the MSD stops to increase. The same conclusion is valid for the stress autocorrelation function:
it rapidly decays if $T>0.15$ and does not decay at all if $T \leq 0.14$. Therefore, the behavior of the dynamical characteristics of
the system is similar in the case of crystallization and QC formation, but qualitatively different in the case of glass transition.

\begin{figure}
\includegraphics[width=8cm,height=8cm]{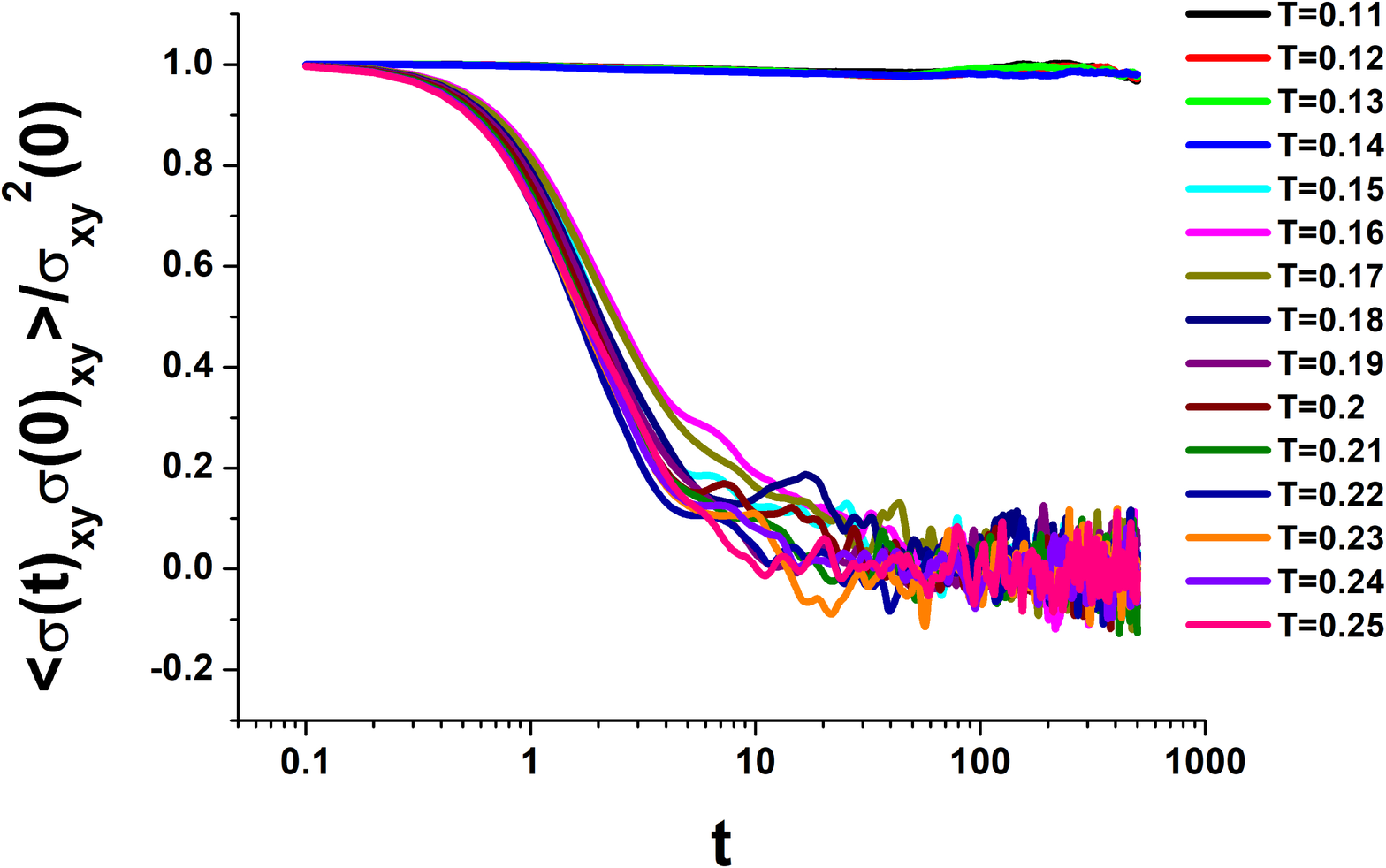}%

\caption{\label{visc-cr} Stress autocorrelation function for RSS with $\sigma_1=1.35$ at $\rho=0.4$.}
\end{figure}

In conclusion, in the present paper we study the behavior of rdfs and several dynamical characteristics of matter upong
crossing the line of crystallization, QC formation and glass transition. We find that the rdfs of QC can be very similar
to the ones of glass and therefore it is not enough to characterize the structure of the system by rdfs only. One needs
more elaborate methods of the structure description. Alternatively, one may compare the dynamical characteristics, like
intermediate scattering function, MSD and stress autocorrelation function. In the case of glass transition
all these functions change smoothly upon cooling. In the case of crystallization and QC formation
they experience a jump, which signalize the formation of a solid phase (crystal or QC). The effect is the most pronounced
in the stress autocorrelation functions. Since the integral of these functions is related to the shear viscosity via
Green-Kubo relation, one can say that in the case of glass transition the viscosity continuously increases with temperature,
while in the case of crystallization of QC formation it remains finite in the liquid phase down to the transition point, while
below it the shear viscosity becomes infinite as it should be in the case of solid.

This work was carried out using computing resources of the federal
collective usage center "Complex for simulation and data
processing for mega-science facilities" at NRC "Kurchatov
Institute", http://ckp.nrcki.ru, and supercomputers at Joint
Supercomputer Center of the Russian Academy of Sciences (JSCC
RAS). The work was supported by the Russian Foundation of Basic Research (Grant No 18-02-00981).


\begin{thebibliography}{99}

\bibitem{kobbinder} K. Binder and W. Kob, Glassy Materials and Disordered Solids, World Scientific Publishing, Singapour (2005).

\bibitem{qc-discovery} D. Shechtman, I. Blech, D. Gratias, J.W. Cahn, Phys. Rev. Lett. 53(20), 1951-1954 (1984).

\bibitem{qc-micelles} St. Fischer, et. al., PNAS 108(5), 1810–1814 (2011).

\bibitem{qc-graphene} W. Yao, et. al., PNAS 115 (27), 6928-6933 (2018).

\bibitem{qc1} M. Dzugutov, Phys. Rev. Lett. 70, 2924 (1993).

\bibitem{qc2} M. Engel, P. F. Damasceno, C. L. Phillips and S. C. Glotzer,
Nat. Mater. 14, 109–116 (2015).

\bibitem{qc3} M. Engel and H.-R. Trebin, Phys. Rev. Lett. 98, 225505 (2007).

\bibitem{qc4} A. S. Keys and Sh. C. Glotzer, Phys. Rev. Lett. 99, 235503 (2007).

\bibitem{qc2d1} A. J. Archer, A. M. Rucklidge, and E. Knobloch, Phys. Rev. Lett. 111, 165501 (2013).

\bibitem{qc2d2} N. P. Kryuchkov,  S. O. Yurchenko, Yu. D. Fomin, E. N. Tsiok, and V. N. Ryzhov, Soft Matter, 14, 2152-2162 (2018).

\bibitem{qc2d3} Yu. D. Fomin, E. A. Gaiduk, E. N. Tsiok, and V. N. Ryzhov, Molecular Physics 116, 3258-3270 (2018).

\bibitem{qc2d4} M. Zu, P. Tan, and N. Xu, Nat. Comm. 8, 2089 (2017).


\bibitem{ryltsev} R. Ryltsev, B. Klumov, and N. Chtchelkatchev, Soft Matter 11, 6991 (2015).

\bibitem{ryltsev1} R. E. Ryltsev, N. M. Chtchelkatchev, Soft Matter 13, 5076-5082 (2017).

\bibitem{ryltsev2} R. E. Ryltsev, Doctoral thesis (in Russian) (2019).

\bibitem{we1} Yu. D. Fomin, N. V. Gribova, V. N. Ryzhov, S. M. Stishov, and D. Frenkel,
J. Chem. Phys. 129, 064512 (2008).

\bibitem{we2} N. V. Gribova, Yu. D. Fomin, D. Frenkel, and V. N. Ryzhov, Phys. Rev. E 79, 051202 (2009).

\bibitem{we3} Yu. D. Fomin, N.V. Gribova and V.N. Ryzhov, DDF 277, 155 (2008).

\bibitem{we4} Yu. D. Fomin, E. N. Tsiok, and V. N. Ryzhov, J. Chem. Phys. 135, 234502 (2011).

\bibitem{we5} Yu. D. Fomin and V. N. Ryzhov, N. V. Gribova, Phys. Rev. E 81, 061201 (2010).

\bibitem{we6} Yu.D. Fomin, V.N. Ryzhov, Physics Letters A 375, 2181–2184 (2011).

\bibitem{we7} Yu.D. Fomin, V.N. Ryzhov, Physics Letters A 377, 1469–1473 (2013).

\bibitem{we8} Yu. D. Fomin, V. N. Ryzhov, B. A. Klumov, and E. N. Tsiok, J. Chem. Phys. 141, 034508 (2014).

\bibitem{we9} Yu. D. Fomin. E. N. Tsiok, V. N. Ryzhov, Phys. Rev. E 87, 042122 (2013).

\bibitem{we10} Yu.D. Fomin, E.N. Tsiok, and V.N. Ryzhov, Eur. Phys. J. Special Topics 216, 165–173 (2013).

\bibitem{we11} Yu. D. Fomin, E. N. Tsiok, and V. N. Ryzhov, J. Chem. Phys. 135, 124512 (2011).

\bibitem{ryltsev-glass} R. E. Ryltsev, N. M. Chtchelkatchev, and V. N. Ryzhov, Phys. Rev. Lett. 110, 025701 (2013).

\bibitem{sn} P. J. Steinhardt, D. R. Nelson and M. Ronchetti, Phys. Rev. Lett., 47, 1297–1300 (1981).

\bibitem{sn1} P. J. Steinhardt, D. R. Nelson and M. Ronchetti, Phys. Rev. B, 28, 784–805 (1983).




\bibitem{lammps} http://lammps.sandia.gov/



\end{thebibliography}
\end{document}